\documentclass[final,1p,times]{elsarticle}

 \usepackage{graphics}
 \usepackage{graphicx}
 \usepackage{epsfig}

 \usepackage{color}
\usepackage{amssymb}
\usepackage{textcomp}
\usepackage{gensymb}
\usepackage{booktabs}
\usepackage{tablefootnote}
\usepackage[nomarkers,nolists]{endfloat}

 \biboptions{sort&compress}

\journal{Journal of Alloys and Compounds}

\begin{document}

\begin{frontmatter}



\title{Magnetic structure of the magnetocaloric compound AlFe$_2$B$_2$}


\author[CHEM]{Johan Cedervall\corref{cor1}}\ead{johan.cedervall@kemi.uu.se}
\author[SSP]{Mikael Svante Andersson}
\author[SSP]{Tapati Sarkar}
\author[PHYS]{Erna K. Delczeg-Czirjak}
\author[KTH]{Lars Bergqvist}
\author[ILL]{Thomas C. Hansen}
\author[REZ]{Premysl Beran}
\author[SSP]{Per Nordblad}
\author[CHEM]{Martin Sahlberg}

\address[CHEM]{Department of Chemistry - \AA{}ngstr\"{o}m Laboratory, Uppsala University, Box 538, 751 21 Uppsala, Sweden.}
\address[SSP]{Department of Engineering Sciences, Uppsala University, Box 534, 751 21 Uppsala, Sweden.}
\address[PHYS]{Department of Physics and Astronomy, Uppsala University, Box 516, 751 20 Uppsala, Sweden}
\address[KTH]{Department of Materials and Nano Physics and Swedish e-Science
Research Centre (SeRC), Royal Institute of Technology (KTH), Electrum 229,
SE-164 40 Kista, Sweden}
\address[ILL]{Institut Laue-Langevin, B.P. 156, 38042 Grenoble Cedex 9, France}
\address[REZ]{Nuclear Physics Institute, Academy of Sciences of the Czech Republic, 25068 Rez, Czech Republic}
\cortext[cor1]{Corresponding author}

\begin{abstract}
The crystal and magnetic structures of AlFe$_2$B$_2$ have been studied with a combination of X-ray and neutron diffraction and electronic structure calculations. The magnetic and magnetocaloric properties have been investigated by magnetisation measurements. The samples have been produced using high temperature synthesis and subsequent heat treatments. The compound crystallises in the orthorhombic crystal system \textit{Cmmm} and it orders ferromagnetically at 285 K through a second order phase transition. At temperatures below the magnetic transition the magnetic moments align along the crystallographic $a$-axis. The magnetic entropy change from 0 to 800 kA/m was found to be -1.3 J/K kg at the magnetic transition temperature. 
\end{abstract}


\end{frontmatter}


\section{Introduction}
\label{}
Since the discovery of the giant magnetocaloric effect (GMCE) in Gd$_5$(Si$_2$Ge$_2$) in 1997 \cite{Pecharsky1997} materials for magnetic refrigeration have been increasingly investigated and discussed. Cooling devices utilising the GMCE are to become 20-30\% more energy efficient than conventional vapour compression systems \cite{GschneidnerJr2008} and could thus contribute to the development of a sustainable global energy economy. The materials showing the highest GMCE contain rare earth elements, which are high in cost and have a negative impact on the environment during mining. Therefore, to increase the sustainability and allow up-scaling, rare earth free materials (both the included magnetocaloric and  permanent magnet material) should be utilised in an application  \cite{Coey2012}. As a consequence there is a need of finding suitable ferromagnetic compounds consisting of cheap and abundant elements. Examples of rare earth free materials for future magnetic cooling devices are FeMnP$_{\rm 1-x}$Si$_{\rm x}$ \cite{Hoglin2015} and the substoichiometric compound Mn$_{1.25}$Fe$_{0.70}$P$_{\rm 1-x}$Si$_{\rm x}$ \cite{Miao2014}, both of which have the hexagonal Fe$_2$P-type structure \cite{Carlsson1973}. These materials have tunable near room temperature Curie temperatures (T$\rm _c$) and high magnetic entropy changes ($\Delta \rm S_{mag}$). 

Another class of potential magnetocaloric materials is AlM$_2$B$_2$ (M = Fe, Mn, Cr) based on AlFe$_2$B$_2$ \cite{Jeitschko1969,ElMassalami2011,Tan2013,Chai2014}. AlFe$_2$B$_2$ crystallises in an orthorhombic structure (space group \textit{Cmmm}) with two formula units per unit cell and cell parameters 2.9233(10), 11.0337(14) and 2.8703(3) $\AA$ for $a$, $b$ and $c$ respectively \cite{Jeitschko1969}. It is a layered structure along the $b$-axis where (Fe$_2$B$_2$)-slabs are alternated with layers of aluminium atoms. In the aluminium layer the atoms occupy the \textit{2a}-position and in the (Fe$_2$B$_2$)-slabs iron and boron occupy the \textit{4j} and \textit{4i} positions, respectively. In previous studies, magnetic measurements and M\"{o}ssbauer spectroscopy have shown that AlFe$_2$B$_2$ is a ferromagnetic material with reported values of T$\rm _c$ varying between 282-320 K and saturation magnetic moments ranging between 0.95 and 1.25 $\mu_B$/Fe-atom \cite{ElMassalami2011,Tan2013,Chai2014}. The magnetic entropy change has been reported to -4.1 and -7.7 J/K kg at an applied magnetic field of 1600 and 4000 kA/m, respectively \cite{Tan2013}.

In this study we report on crystallographic and magnetic properties of AlFe$_2$B$_2$, including the magnetic structure, investigated using XRD, neutron diffraction, magnetic measurements as well as electronic structure calculations. Our study reveals the low temperature magnetic structure of AlFe$_2$B$_2$, reports the behaviour of the lattice parameters through the temperature region of the ferromagnetic transition and suggests that this transition is of second order. 

\section{Experiments}
\label{}

\subsection{Sample preparation}
The samples were synthesised by arc melting stoichiometric amounts of iron (Leico Industries, purity 99.995\%. Surface oxides were reduced in H$_2$-gas.), boron (Wacher-Chemie, purity 99.995\% or Los Alamos National Laboratories, pure $^{11}$B) and aluminium (Gr\"anges SM, purity 99.999\%) under an argon atmosphere. To suppress formation of unwanted phases; FeB was first prepared and then AlFe$_2$B$_2$ was produced with a 50\% excess amount of  aluminium \cite{ElMassalami2011}. To ensure maximum homogeneity the samples were remelted 5 times in total. The samples were crushed and pressed into pellets, placed in evacuated silica tubes, annealed for 14 days at 1173 K and subsequently quenched in cold water. Some additional phases were removed by etching the samples in diluted HCl (1:1) for 10 minutes. Two samples were prepared, one using natural boron and the other using isotopically pure $^{11}$B to allow neutron diffraction experiments.

\subsection{X-ray powder diffraction}
To analyse the crystalline phases and to perform crystal structure determinations X-ray powder diffraction (XRD) were performed with a Bruker D8 diffractometer equipped with a Lynx-eye position sensitive detector (PSD, 4\textdegree\ opening) using CuK$\alpha_1$ radiation ($\lambda$ = 1.540598 \AA ), in a 2$\theta$ range of 20-90\textdegree\ at room temperature. Temperature dependant (16-295 K) diffraction patterns were also recorded to study changes in the unit cell parameters through the magnetic transition with added Si as an internal standard.

\subsection{Neutron powder diffraction}
The nuclear and magnetic structures were determined by refinement of neutron powder diffraction patterns collected at the D1B beamline at ILL (Grenoble, France). A pyrolytic graphite monochromator (reflection 002) was used, giving a neutron wavelength of 2.52 \AA. Diffraction patterns were recorded within a 2$\theta$ range of 5-128\textdegree, at the discrete temperatures 16, 150, 200, 300 and 500 K as well as upon ramping between these temperatures \cite{ExpdataILL}. The sample was kept in a double-walled V container to minimise absorption by the sample.

\subsection{Refinement of the crystal and magnetic structures}
The X-ray and neutron powder diffraction patterns were analysed with the FullProf software \cite{Rodriguez-Carvajal1993} utilising the Rietveld method \cite{Rietveld1969}. The pseudo-Voigt profile function was used to describe the profile function and the background was described by linear interpolation between chosen points. In the refinement the following 16 parameters were varied: zero point, background, scale factor, peak shape, half width parameters (3), unit cell parameters (3), atom occupancies (3), isotropic temperature parameter and atomic coordinates (2). To precisely determine the unit cell parameters refinements were performed in the software UnitCell \cite{Holland1997}. 

Symmetry analysis of possible magnetic arrangement were done using the software package Sarah \cite{Wills2000}. Different irreducible representation were tested against measured neutron diffraction patterns. 

\subsection{Magnetisation}
Magnetic measurements were made using an MPMS SQUID magnetometer from Quantum design as well as a PPMS VSM system also from Quantum Design. Magnetisation (M) as a function of temperature (T) measurements were carried out in the temperature interval 20 to 400 K using both Field Cool Cooling (FCC) and Field Cool Warming (FCW) protocols. The M vs. T measurements were made using two different fields 4 and 800 kA/m. Magnetisation as a function of magnetic field (H) measurements were made at different temperatures, 10 K and 260 to 305 K in 2.5 K steps. The series of measurements from 260 to 305 K were made to estimate the magnetic entropy $\Delta \rm S_{mag}$ using the equation \begin{equation}
\Delta S_{mag}=\mu_0 \int^{H_f}_{H_i}{\left(\frac{dM}{dT}\right)_H}dH 
\label{eq:deltaS}
\end{equation} where $\mu_0$ is the magnetic permeability of vacuum, and $H_i$ and $H_f$ are the lowest and highest values of the magnetic field \cite{deOliveira2008}. The magnetic data from 260 to 305 K were also used to construct an Arrot plot, to determine the order of the phase transition. \cite{Arrott1957,Arrott1967,Banerjee1964}.

\subsection{Electronic structure calculations}
Electronic structure calculations were carried out within the density functional theory framework to investigate the electronic structure and magnetic properties of AlFe$_2$B$_2$ using spin polarized relativistic Korringa-Kohn-Rostoker (SPR-KKR) method \cite{spr-kkr}. The Perdew-Burke-Ernzerhof exchange-correlation approximation \cite{Perdew1996} gave a good description of the lattice parameters and magnetic moments. The paramagnetic phase was modeled within the coherent
potential approximation \cite{Soven1967} combined with the disordered local moment \cite{Pindor1983, Staunton1984} approach. This approximation accurately describes the paramagnetic state with randomly oriented local magnetic moments
\cite{Pindor1983}. Here, the AlFe$_2$B$_2$ ternary system was treated as a quaternary Al(Fe$^{\uparrow}_{0.5}$Fe$^{\downarrow}_{0.5}$)$_2$B$_2$ alloy, with a random mixture of two magnetic states of Fe. The orientation of the magnetic moments have been evaluated as a difference of the total energies calculated for the three magnetisation directions, along $a$, $b$ and $c$ axes, respectively, without considering the full-potential effects. The basis set consisted of $s$, $p$, $d$ and $f$ orbitals ($l_{\rm max}$=3) and the
number of {\bf k} points was set to 340, 2000 and 11000 for the calculation of the ground state properties, density of states (and magnetic exchange integrals) and orientations of the magnetic moments, respectively.

\section{Results and discussion}
\label{}

\subsection{Phase analysis}
\begin{figure}[t!]
  \centering
\includegraphics[width = 1.00\textwidth]{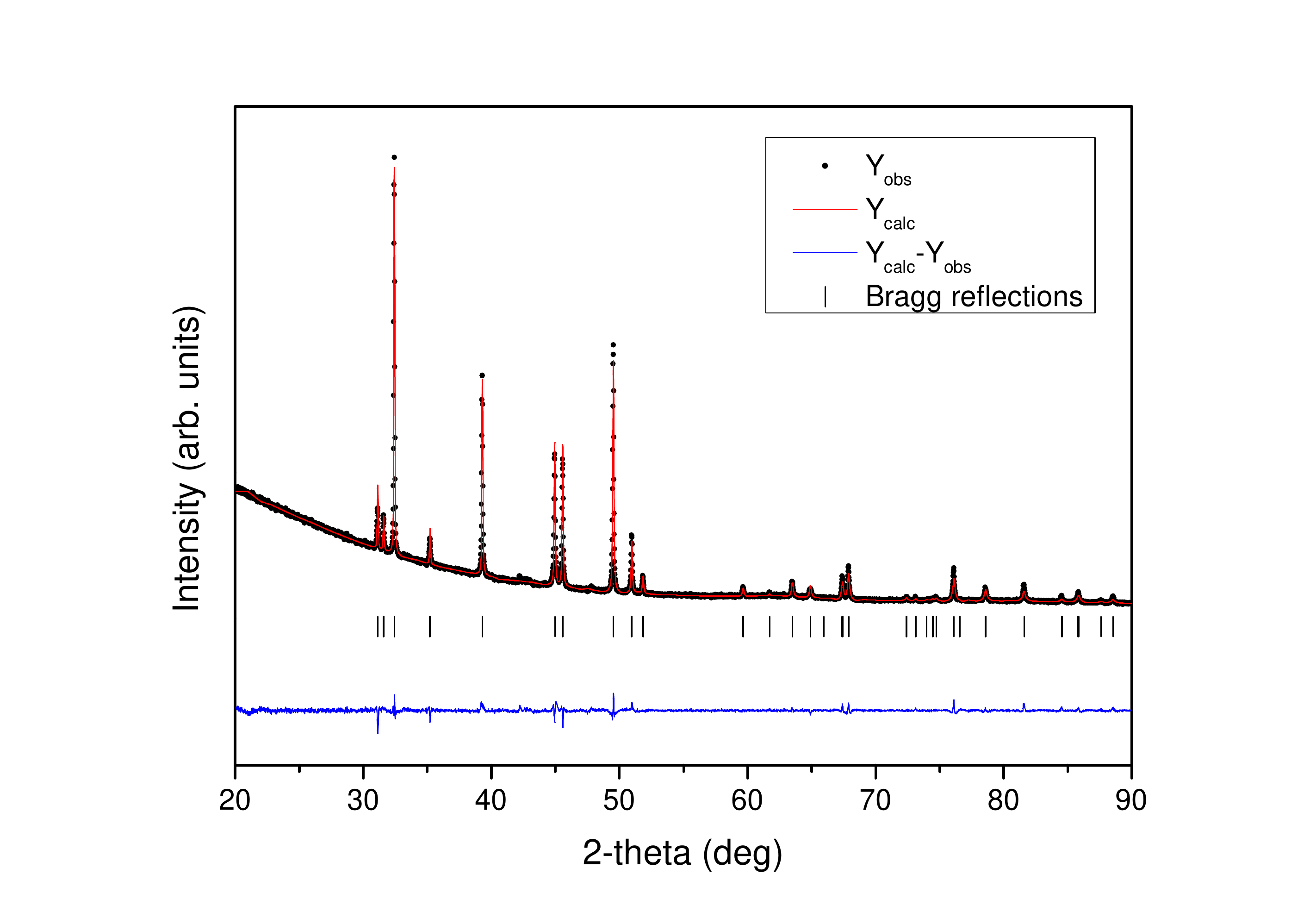}
\caption{X-ray powder diffraction pattern of AlFe$_2$B$_2$ refined with the Rietveld method. Black dots and red line corresponds to the observed and calculated pattern respectively. Blue line shows the difference between observed and calculated data. Tick marks indicate the theoretical Bragg peaks of AlFe$_2$B$_2$.}
\label{fig-x-ray}
\end{figure}

The recorded and calculated intensities of the XRD patterns at room temperature for the sample prepared from natural boron are shown in figure \ref{fig-x-ray} confirming that AlFe$_2$B$_2$ crystallises in the orthorhombic space group \textit{Cmmm} with the unit cell parameters $a$ = 2.9256(4) \AA, $b$ = 11.0247(4) \AA{} and $c$ = 2.8709(2) \AA{}. The atomic coordinates and occupancies obtained from the refined structure model are listed in table \ref{tab:atom-pos} and are in agreement with results from previous studies \cite{Jeitschko1969}.

\begin{table*}[t!]
\small
\caption{Atomic coordinates and occupancies for AlFe$_2$B$_2$ at room temperature from XRD refinement.}
\begin{tabular*}{\textwidth}{@{\extracolsep{\fill}}llllll}
\toprule
Atom	& Site	& x 				& y					& z					& Occ.			\\
\midrule
Fe		& 4j		& 0					& 0.3543(1)	& 0.5				& 0.250(1)	\\
B			& 4i		& 0					& 0.2112(1)	& 0					& 0.250(1)	\\
Al		& 2a		& 0					& 0					& 0					& 0.125(1)	\\
\midrule
\multicolumn{6}{l}{R$_{Bragg}$ = 3.47, R$_{wp}$ = 14.8}				\\
\bottomrule
\label{tab:atom-pos}
\end{tabular*}
\end{table*}

\begin{figure}[t!]
  \centering
\includegraphics[width = 1.00\textwidth]{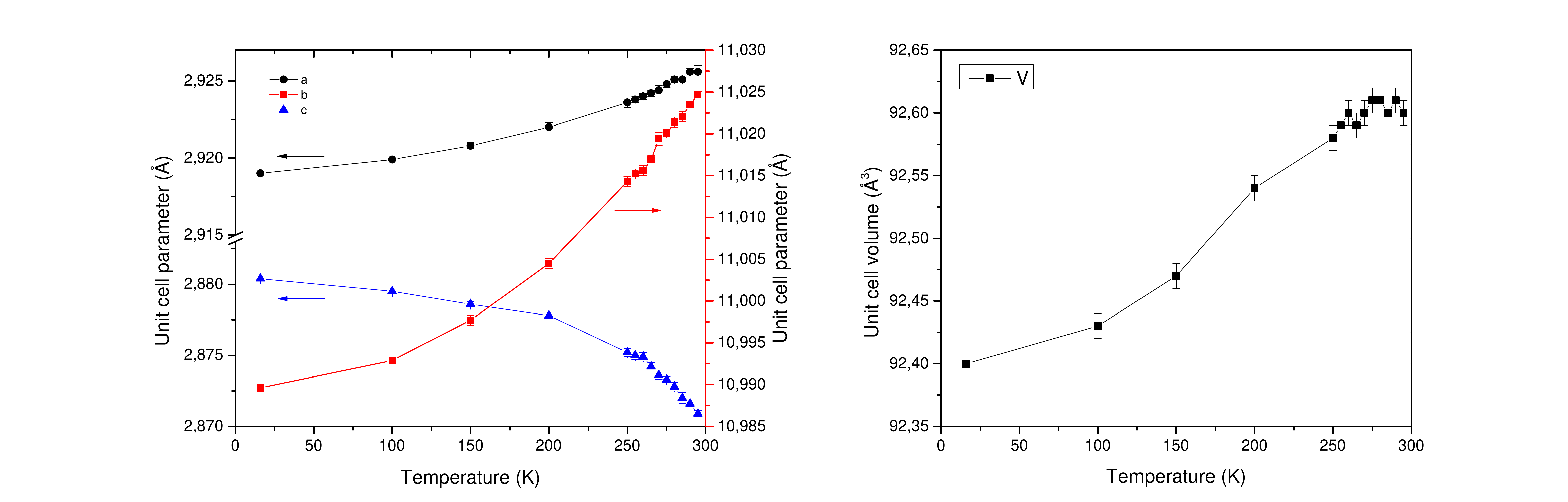}
\caption{The development of the unit cell parameters (left) and unit cell volume (right) as a function of temperature for AlFe$_2$B$_2$. The vertical dashed line marks T$\rm _c$.}
\label{fig-lowtemp-unitcell}
\end{figure}

The unit cell parameters were refined for the collected temperature dependant XRD patterns of the AlFe$_2$$^{11}$B$_2$ sample and the results are presented in figure \ref{fig-lowtemp-unitcell}. As can be seen, the $a$ and $b$ parameters decrease continuously while $c$ increases with decreasing temperature. The different changes of the unit cell parameters give a rather constant unit cell volume (right panel of figure \ref{fig-lowtemp-unitcell}) down to 250 K where the cell volume starts to decrease with decreasing temperature. However, the discontinuities previously reported to occur in the temperature dependence of the lattice parameters near the transition temperature \cite{Lewis2015} are not reproduced in these experiments.

Most of the experiments in this investigation were performed on the sample that was synthesised with pure $^{11}$B to allow neutron diffraction measurements. This sample was not as phase pure as the sample made with natural boron, e.g. the occupancies of Fe is slightly lower in the $^{11}$B-sample compared to the one with natural boron and there is existence of non-magnetic secondary phases. However, for consistency with the results from the neutron diffraction studies, the results from the $^{11}$B-sample are mainly reported, but results from the natural boron sample are included where comparison is instructive.

\subsection{Magnetisation and magnetocaloric effect}
\begin{figure}[t!]
  \centering
\includegraphics[width = 0.6666\textwidth]{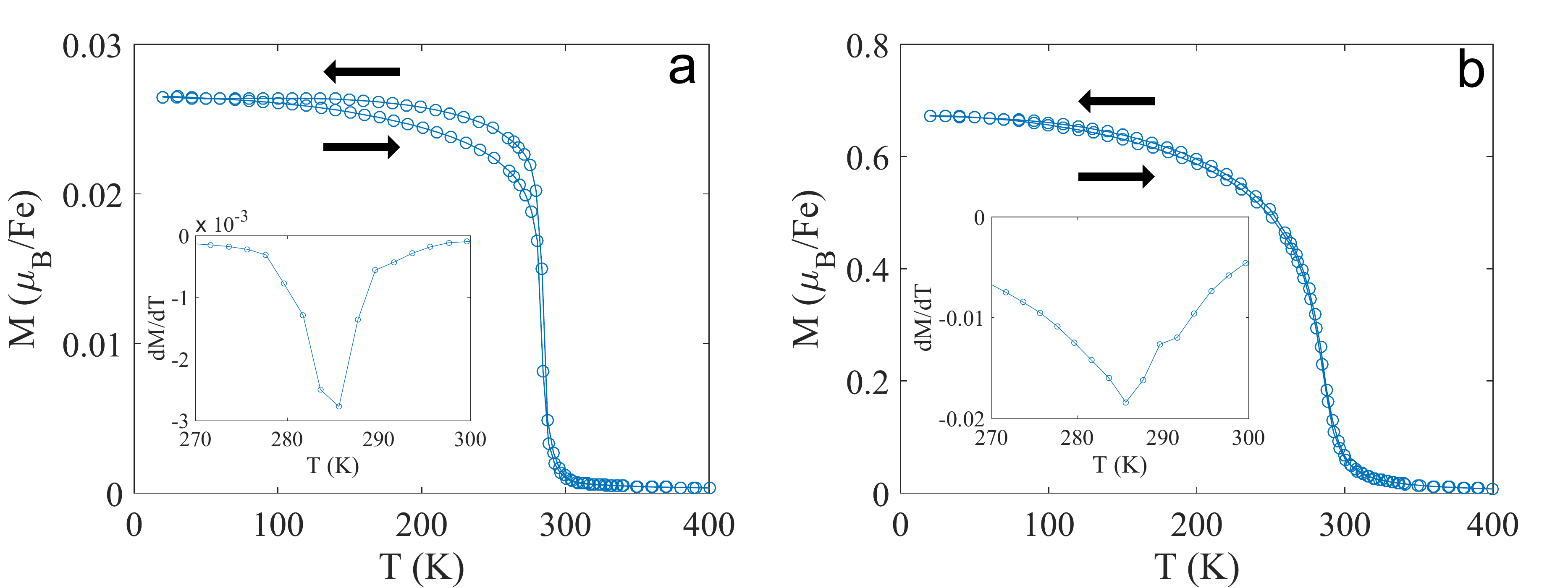}
\caption{FCC and FCW magnetisation as a function of temperature using a field of (a) 4 kA/m and (b) 800kA/m. The insets show the temperature derivative of the FCC magnetisation.}
\label{fig-FCC_FCW}
\end{figure}

\begin{figure}[t!]
  \centering
\includegraphics[width = 0.50\textwidth]{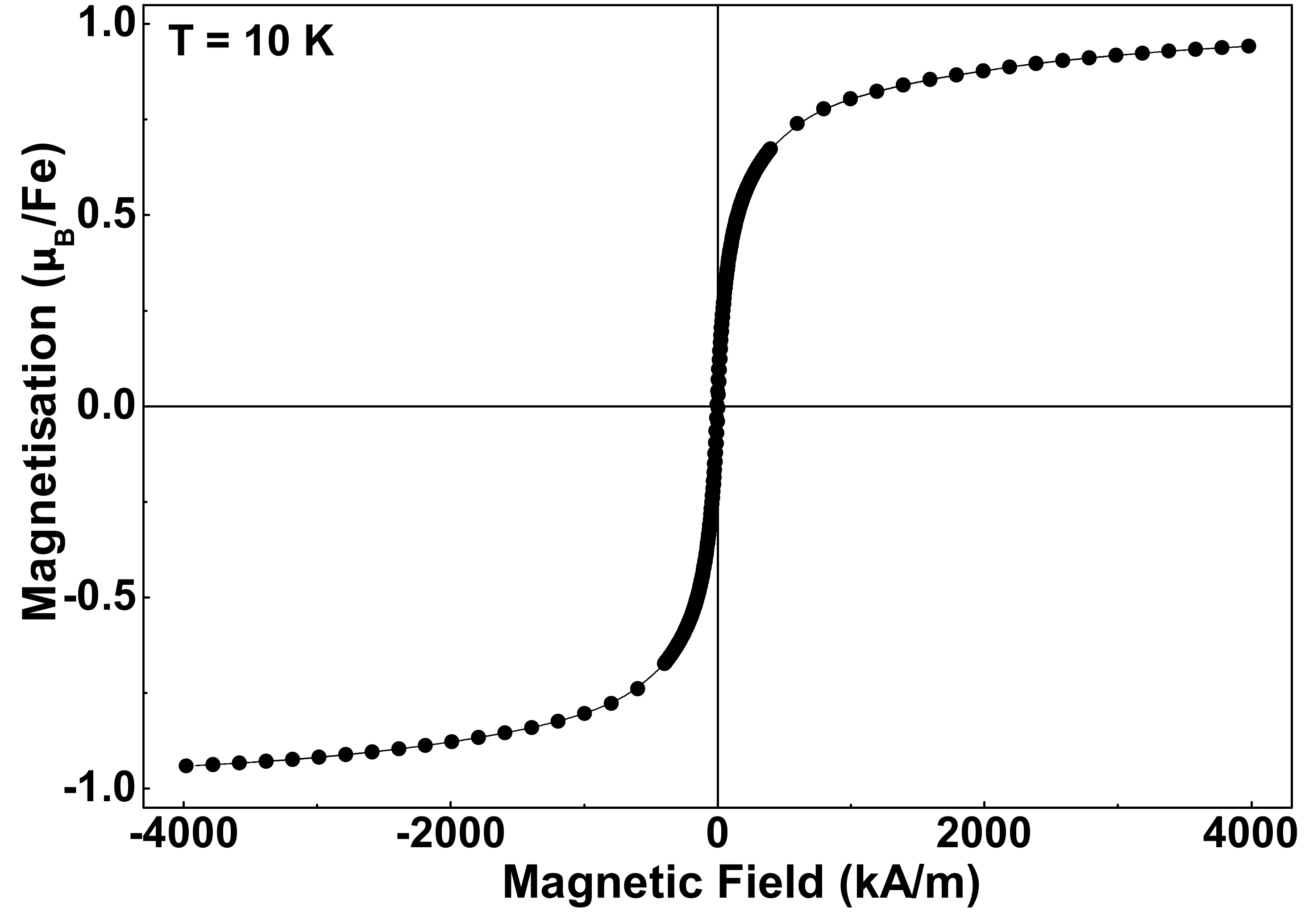}
\caption{Magnetisation vs. magnetic field of AlFe$_2$B$_2$ recorded at T = 10 K.}
\label{fig-MvsHat10K}
\end{figure}

Figure \ref{fig-FCC_FCW} shows the magnetisation vs. temperature data recorded in the FCC and FCW modes under two different magnetic fields, 4 kA/m (figure \ref{fig-FCC_FCW}a) and 800 kA/m (figure \ref{fig-FCC_FCW}b). The ferromagnetic transition temperature can be estimated from the derivative of the M vs. T curves. From the insets of figure \ref{fig-FCC_FCW}, T$\rm _c$ can be extracted to be 285 K. There is no appreciable thermal hysteresis between the cooling and warming curves near the transition temperature. This is an indication that, unlike a recent report of a first order transition in AlFe$_2$B$_2$ \cite{Lewis2015}, the sample shows a second order magnetic transition. It should here be mentioned that the measurement protocols employed in this study have been done using the settle mode (the temperature is stabilised at each point before measuring the magnetisation) so as to eliminate any spurious effects due to temperature lag which often occur when measurements are made in sweep mode. In figure \ref{fig-MvsHat10K} the M vs. H curve recorded at 10 K is shown. The saturation magnetisation of the sample, estimated from magnetisation value at 4000 kA/m, turns out to be 0.9 $\mu_B$/Fe, which is lower than the moment derived from our theoretical (1.1 $\mu_B$/Fe) as well as  neutron diffraction studies (1.4 $\mu_B$/Fe). The presence of non-magnetic impurity phases in the studied sample could well account for this difference. The AlFe$_2$B$_2$ sample made from the natural isotopical composition of boron was also measured, yielding a somewhat higher T$\rm _c$ (295 K) and higher saturation magnetisation (1.0 $\mu_B$/Fe).

\begin{figure}[t!]
  \centering
\includegraphics[width = 1.00\textwidth]{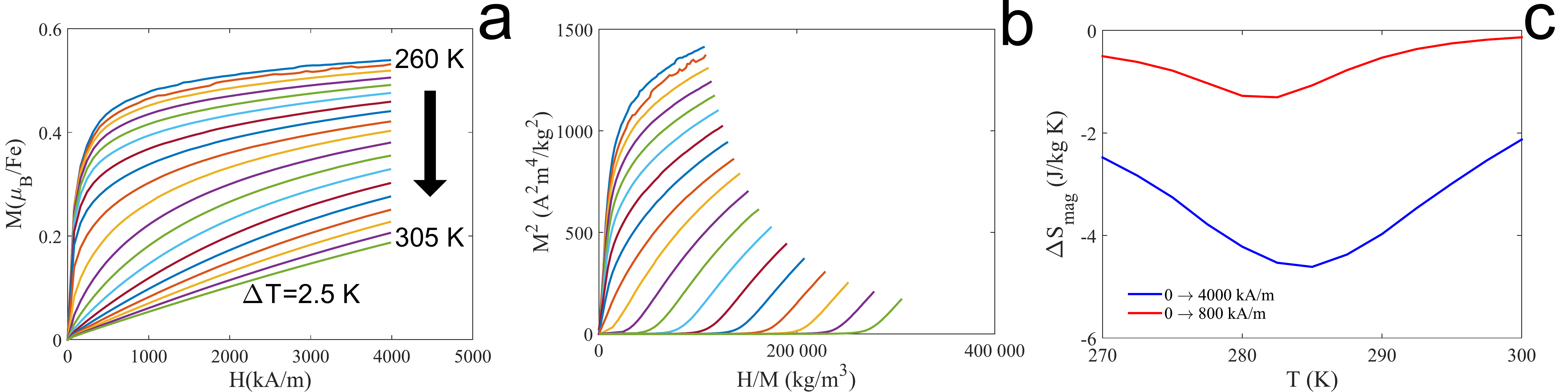}
\caption{ (a) Magnetisation as a function of magnetic field at temperatures between 260 and 305 K (steps of 2.5 K). (b) An Arrot plot of the data shown in (a). (c) $\Delta \rm S_{mag}$ as function of temperature estimated using the data shown in (a) for a field change of 0 to 800 kA/m and 0 to 4000 kA/m.}
\label{fig-MvsH_deltaS_Arrot}
\end{figure}

The M vs. H curves at several temperatures around the transition temperature (T ranging from 260 K to 305 K, with temperature intervals of 2.5 K) are shown in figure \ref{fig-MvsH_deltaS_Arrot}. This measurement serves a dual purpose. Firstly, the possibility to check the order of the transition. The nature of the magnetic transition can be obtained from the slope of isotherm plots of M$^2$ vs. H/M of this Arrot plot. This criterion for deciding the nature of the transition is generally referred to as the Banerjee criterion \cite{Banerjee1964}, and has been widely used to experimentally determine the order of the magnetic phase transition. Briefly, a positive or a negative slope of the experimental M$^2$ vs. H/M curve indicates a second order or a first order transition, respectively. The Banerjee criterion provides a straightforward way to distinguish between samples showing a first order transition and those showing a second order transition, because the slope of the M$^2$ vs. H/M curves changes sign. This criterion have been applied in this investigation (figure \ref{fig-MvsH_deltaS_Arrot}b). As is clear from the figure, the slope of the M$^2$ vs. H/M curves are positive throughout the measured temperature range, confirming that the sample shows a second order transition. 

The curves shown in figure \ref{fig-MvsH_deltaS_Arrot}a have also been used to estimate the magnetic entropy change ($\Delta \rm S_{mag}$) in the sample. The plot of $\Delta \rm S_{mag}$ vs. temperature is shown in figure \ref{fig-MvsH_deltaS_Arrot}c. The maximum entropy change occurs at around T$\rm _c$ = 285 K, and is -1.3 J/kg K for a change of magnetic field from 0 to 800 kA/m, and -4.5 J/kg K for a change of magnetic field from 0 to 4000 kA/m.

\subsection{Magnetic structure}
\begin{figure}[t!]
  \centering
\includegraphics[width = 1.00\textwidth]{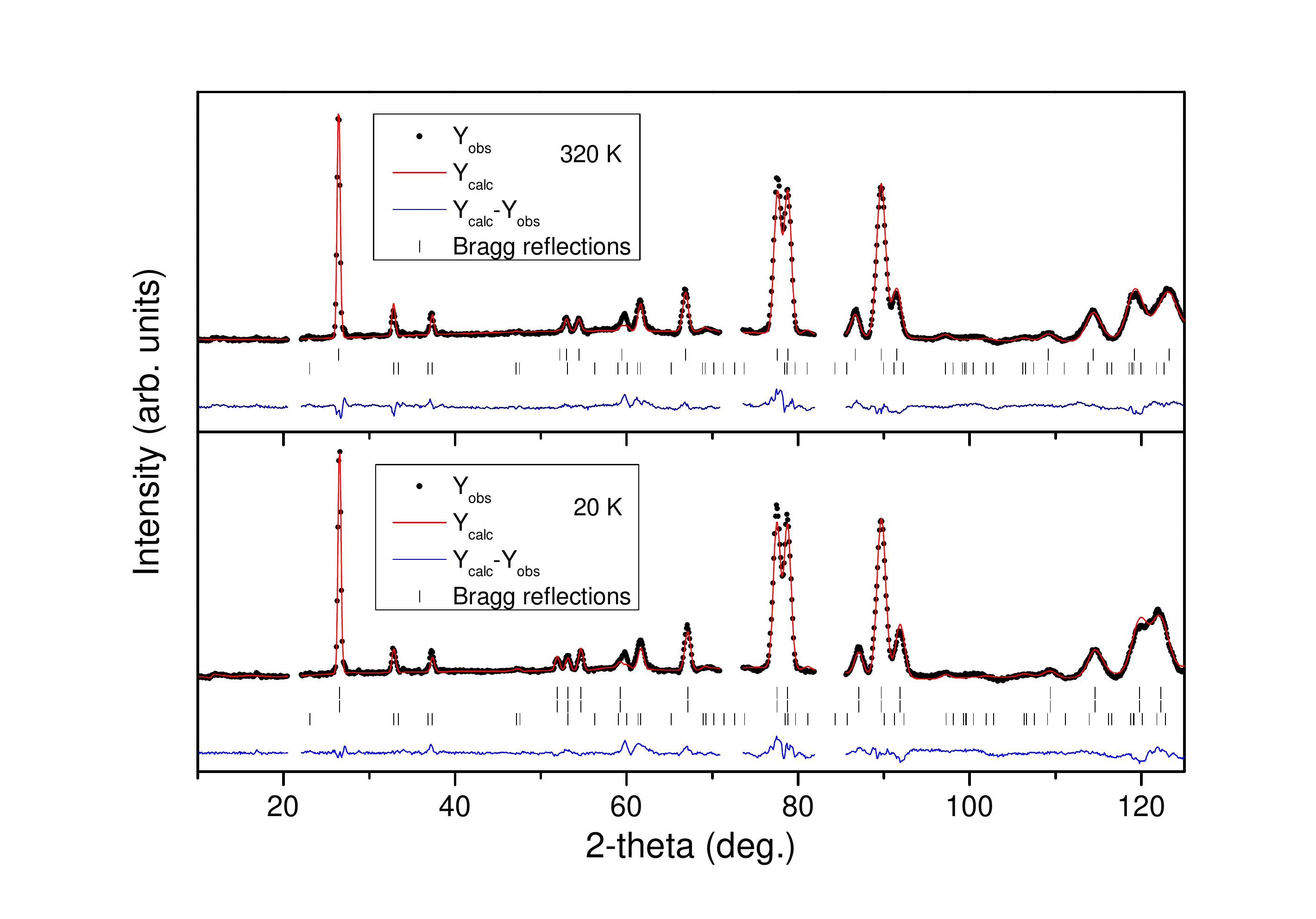}
\caption{Neutron powder diffraction patterns of AlFe$_2$B$_2$ refined with the Rietveld method at 320 K (upper) and 20 K (lower). The tick marks indicate the position of the Bragg reflections for the phases AlFe$_2$B$_2$ (upper) and B (lower); at 20 K tick marks for a magnetic AlFe$_2$B$_2$ phase are also included. $\lambda$ = 2.52 \AA.}
\label{NPD-refined}
\end{figure}

The measured and calculated intensities fron neutron diffraction at 320 K and 20 K are presented in figure \ref{NPD-refined}. As seen in the powder diffractograms, there are additional phases present in the $^{11}$B-sample. Most of the peaks that does not belong to the main phase ($\sim$5 \%) can be explained by small amounts of tetragonal boron ($a$ = 8.917(2) \AA, $c$ = 5.025(7) \AA, with the space group \textit{P$\bar{4}$n2} \cite{Hoard1951}). A second additional phase could be identified using a orthorhombic unit cell ($a$ =17.081(7) \AA, $b$ = 11.354(8) \AA, $c$ = 2.288(1) \AA, space group = \textit{Pnma}). This second additional phase could not be identified even though all reported binary and ternary phases in the Al-Fe-B system were tested, as well as their oxides, hydroxides and chlorides. 

\begin{figure}[t!]
  \centering
\includegraphics[width = 1.00\textwidth]{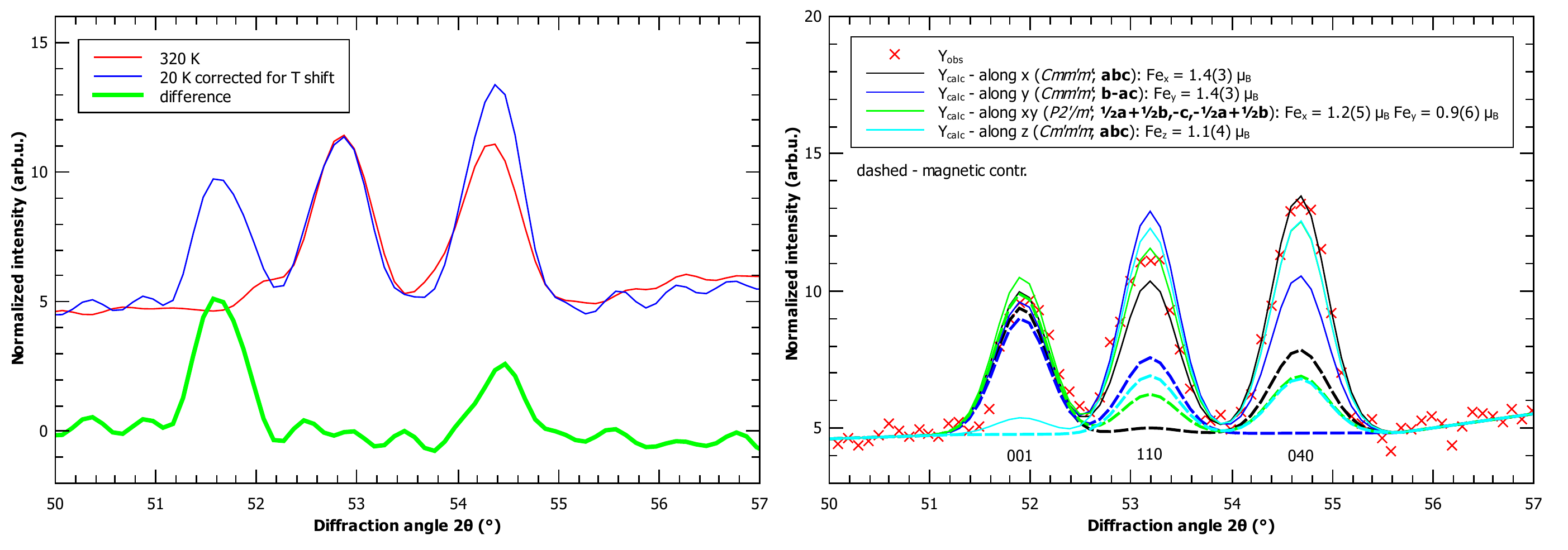}
\caption{Difference curve for the temperatures 320 K and 20 K, the 20 K dataset has been shifted by -0.2\textdegree{} to compensate for thermal expansion effects (a) and a comparison of the refinements for the different magnetic models (b). $\lambda$ = 2.52 \AA.}
\label{NPD-difference}
\end{figure}

In the difference plot, figure \ref{NPD-difference}a, the clear intensity differences occurs for the (001) and (040) reflections at 52.0\textdegree\ and 54.8\textdegree\ in 2$\theta$. Difference in the intensities of these reflections can be explained by magnetic contribution with the same magnetic unit cell as the nuclear phase (propagation vector k = (000)). In order to determine the magnetic structure of AlFe$_2$B$_2$ several symmetry allowed magnetic structure models were tried in which the moments were aligned along the different crystallographic axis, as well as more complex models with e.g. moments aligned in the $ab$-plane. The contribution of the different magnetic models to the calculated total intensities in comparison with measured ones are shown in figure \ref{NPD-difference}b.

\begin{figure}[t!]
  \centering
\includegraphics[width = 1.00\textwidth]{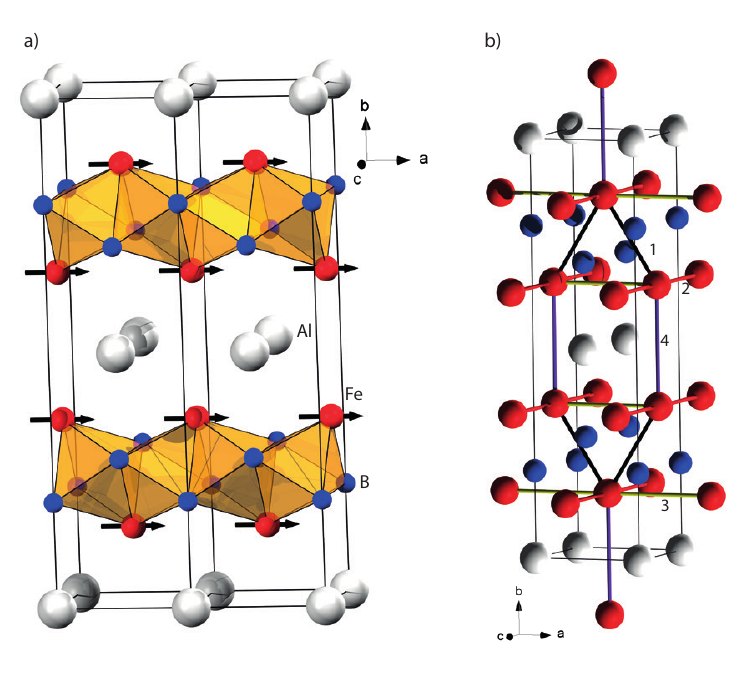}
\caption{Two unit cells of AlFe$_2$B$_2$ with arrows indicating the magnetic structure at 20 K are shown in (a). (b) shows the first four nearest neighbours in AlFe$_2$B$_2$. Bonds between 1$^{\rm st}$, 2$^{\rm nd}$, 3$^{\rm rd}$ and 4$^{\rm th}$ NN are labelled by black, dark red, yellow and purple colours, respectively.}
\label{Mag-structure}
\end{figure}

By comparing the different magnetic models, the best agreement with the observed intensities is achieved by aligning the magnetic moments ferromagnetically along the $a$-axis of the AlFe$_2$B$_2$ unit cell, with the Shubnikov Group \textit{Cmm'm'}, refinement parameters are given in table \ref{tab:mag}. This gives a magnetic moment of 1.4(3) $\mu_B$/Fe-atom. All other models give magnetic intensities larger than 0 for the (110) peak in combination with under- or overestimating the (001) and (040) peaks in the diffractogram (figure \ref{NPD-difference}b), something which is not in agreement with the observed difference in intensities between temperatures. This model is also in agreement with the results from the performed ab initio calculations. The proposed model can be seen in figure \ref{Mag-structure}a.

\begin{table*}[t!]
\small
\caption{Placement, occupancies and magnetic moments of the magnetic Fe-atoms for 320 and 20 K from neutron diffraction refinements.}
\begin{tabular*}{\textwidth}{@{\extracolsep{\fill}}llllllllll}
\toprule
Atom	& Site	& \multicolumn{4}{l}{320 K} & \multicolumn{4}{l}{20 K} \\
\cline{3-10}
 & & x & y & z & Occ. & x & y & z & Magnetic moment ($\mu_B$) \\
\midrule
Fe		& 4j		& 0			& 0.3549(1)	& 0.5	& 0.250(1) & 0 & 0.3560(1)	& 0.5	& 1.4(3)	\\
B			& 4i		& 0			& 0.2075(1)	& 0		& 0.250(1) & 0 & 0.2065(1)	& 0		& -	\\
Al		& 2a		& 0			& 0					& 0		& 0.125(1) & 0 & 0					& 0		& - \\
\midrule
 & & \multicolumn{4}{l}{R$_{Bragg}$ = 7.23, R$_{wp}$ = 12.1} & \multicolumn{4}{l}{R$_{Bragg}$ = 6.30, R$_{mag}$ = 5.16, R$_{wp}$ = 13.8} \\
\bottomrule
\label{tab:mag}
\end{tabular*}
\end{table*}

This model is similar to that of the low temperature magnetic structure of Fe$_5$SiB$_2$ \cite{Cedervall-Fe5SiB2}, which has a layered structure with magnetic moments aligned along the Fe-B-layer in the structure. A recent M\"ossbauer study on a textured sample ($b$-axis out of the plane) of AlFe$_2$B$_2$ suggests that the magnetic moments are in the $ab$-plane, with an angle of \textless 40\textdegree\ from the $b$-axis \cite{AlFe2B2-Mossbauer}. That there is no magnetic order along the $c$-direction is in support of the model presented in this paper, but the neutron diffraction data does not show any magnetic contributions indicating a $b$-component.

\subsection{Electronic structure}
The calculated equilibrium lattice parameters and volumes are listed in table  \ref{theory_table} in comparison with the experimental data for ferromagnetic (FM) and paramagnetic (PM) phases of AlFe$_2$B$_2$. An excellent agreement between the calculated and measured values were found.

\begin{table*}[t!]
\small
\caption{Theoretical and experimental lattice parameters ($a$, $b$, $c$) and volumes ($V$) for FM and PM AlFe$_2$B$_2$. $\delta_{\rm FM}$ ($\delta_{\rm PM}$) represents the deviation of the theoretical crystal parameters calculated for FM (PM) phase and the experimental values measured at 16 (300) K.}
\begin{tabular*}{\textwidth}{@{\extracolsep{\fill}}llllll}
	&	& $a$($\AA$)	& $b$($\AA$)	& $c$($\AA$)	& $V$($\AA^3$)	\\
\midrule
FM	& Theoretical & 2.917			& 11.010			& 2.864			& 91.972		\\
		& Experimental & 2.9190(1) & 10.9896(3) & 2.8804(1) & 92.40(1)	\\
		& $\delta_{\rm FM}$ & -0.1\%	& 0.2\%		& -0.6\%		& -0.5\%		\\
\midrule
PM	& Theoretical & 2.917			& 11.011			& 2.864			& 92.001		\\
		& Experimental & 2.9256(4) & 11.0247(4) & 2.8709(2) & 92.60(1)	\\
		& $\delta_{\rm PM}$	& -0.2\%	& -0.2\%	& -0.1\%		& -0.5\%		\\
\bottomrule
\label{theory_table}
\end{tabular*}
\end{table*}

\begin{figure}[t!]
 \centering
\includegraphics[width=75mm]{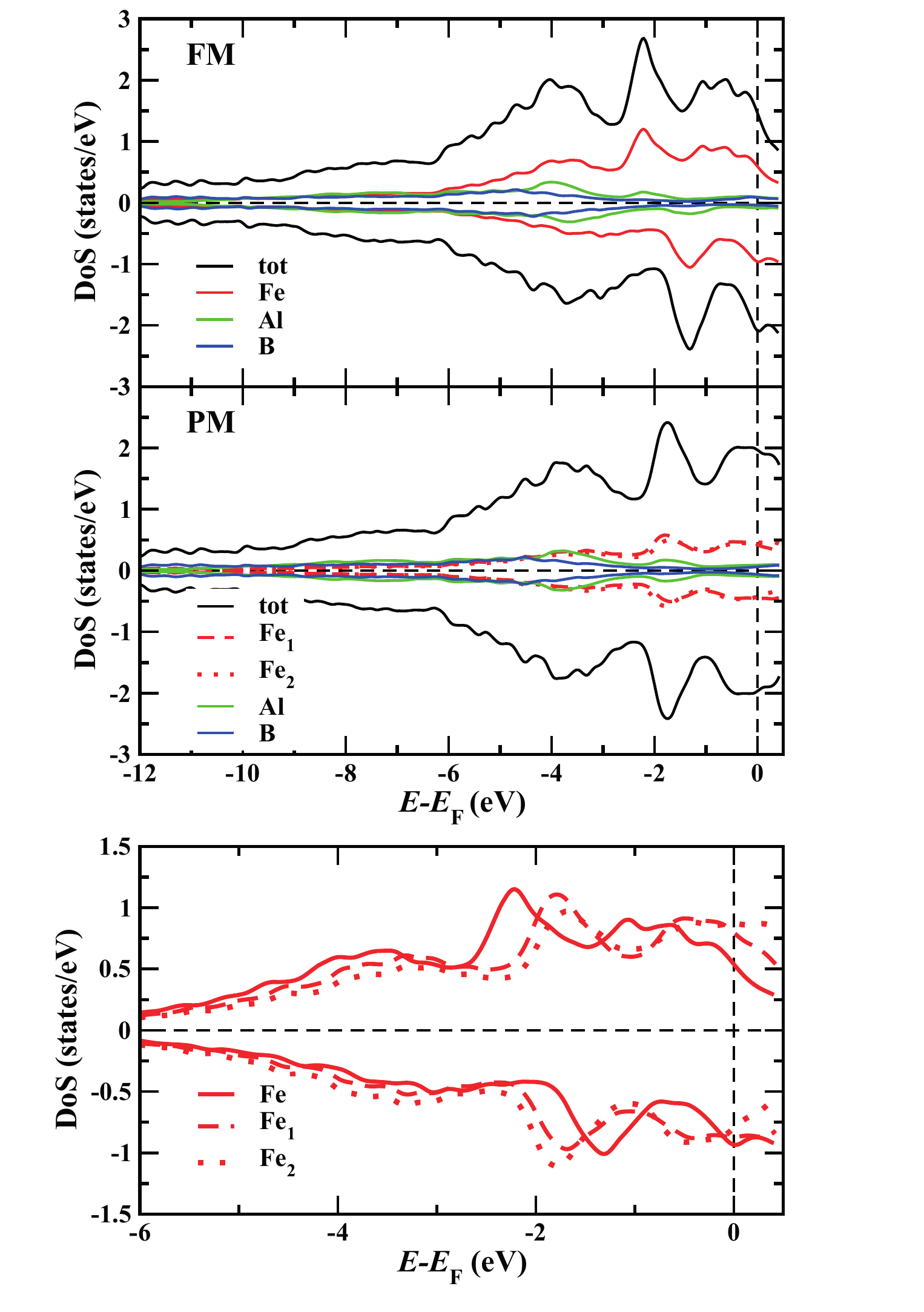}
\caption{Spin and site projected densities of states of FM (upper panel) and PM (middle panel) AlFe$_2$B$_2$. Orbital projected Fe densities of states are plotted for FM (labeled as Fe, full line) and PM (Fe$^\uparrow$, dashed
line and Fe$^\downarrow$, dotted line) AlFe$_2$B$_2$ in the lower panel.}
\label{dosfig}
\end{figure}

The spin polarized densities of states (DOS) for FM and PM phases of AlFe$_2$B$_2$ is presented in the upper and middle panel of figure \ref{dosfig}. In the FM phase, AlFe$_2$B$_2$ shows a spin-polarization that results in a magnetic moment of 1.077 $\mu_{\rm B}$/Fe-atom. Very small magnetic moments are induced on Al and B, -0.05 and -0.02 $\mu_{\rm B}$, respectively, and they couple anti-ferromagnetically to the Fe atoms. The total magnetisation for the modelled PM configuration is zero. As is presented in the middle panel of figure \ref{dosfig} the spin-up and spin-down channel cancel each other. However, there is an observable shift between the spin-up and spin-down channel of the Fe 3$d$ DOS (see lower panel of figure \ref{dosfig}), resulting in 0.3841 $\mu_{\rm B}$ local moment of Fe. The existence of local Fe moments in the modelled paramagnetic phase indicates that a metamagnetic phase transition is not likely in case of AlFe$_2$B$_2$.   

\begin{figure}[t!]
 \centering
\includegraphics[width=75mm]{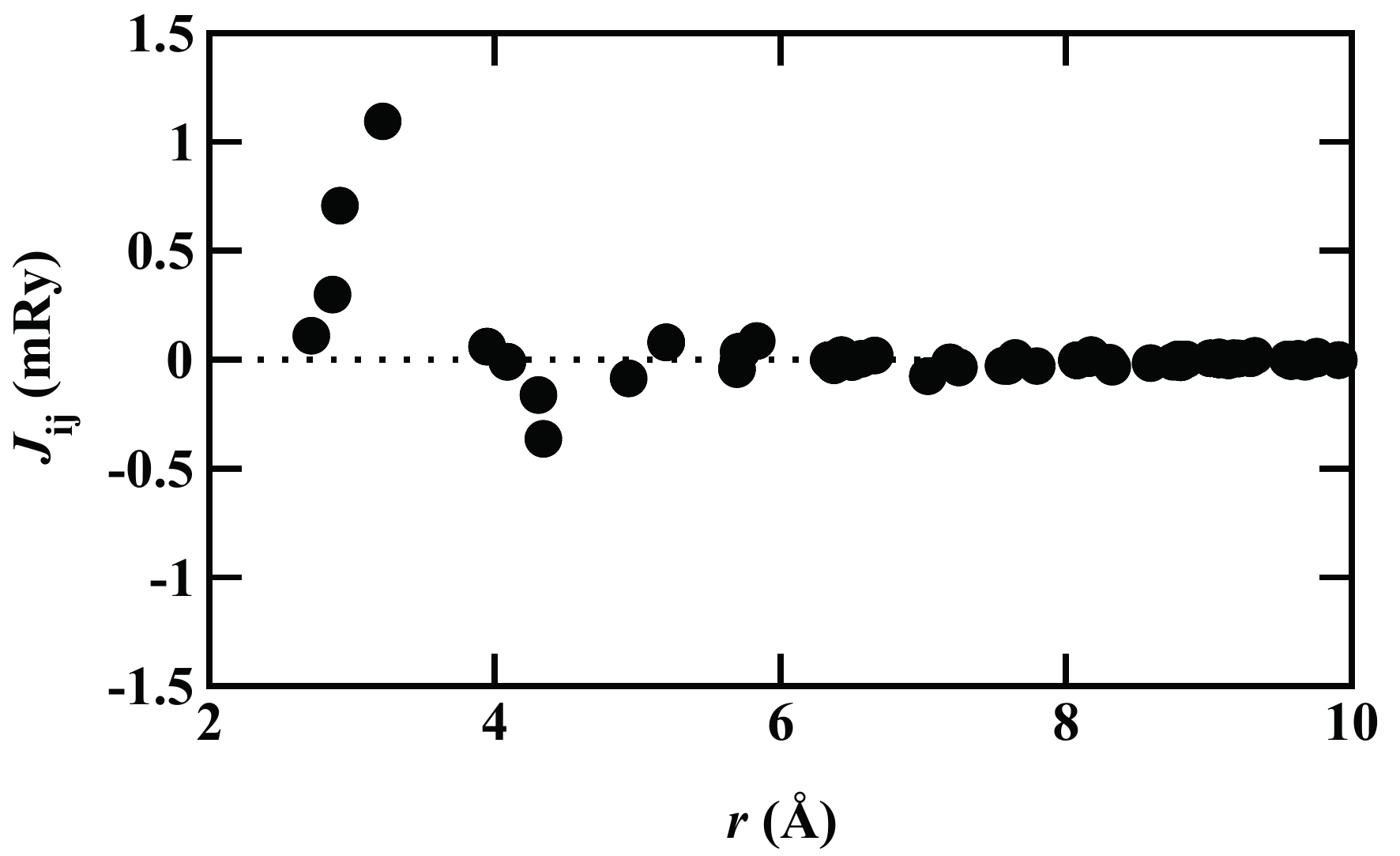}
\caption{Exchange integrals as a function of distance calculated for FM AlFe$_2$B$_2$.}
\label{jijfig}
\end{figure}

Exchange integrals calculated for the FM configuration of AlFe$_2$B$_2$ are plotted in figure \ref{jijfig} as a function of distance. An increasing trend for the first four interactions is observed. These are $J_{\rm ij}$'s between the 1st four Fe nearest neighbours (NN) shown in figure \ref{Mag-structure}b. The 1$^{\rm st}$ (black bonds) and 3$^{\rm rd}$ (yellow bonds) interactions are between Fe atoms situated in a plane that lies in between B-Al layers. The 2$^{\rm nd}$ (dark red bonds) and 4$^{\rm th}$ NN $J_{\rm ij}$ (purple bonds) describes the interaction between Fe atoms separated by B-Al layer and Al layer, respectively, the latter interaction
being the largest one. These findings indicate that substitution of the non-magnetic elements may crucially affect the strength of the magnetic interaction between the Fe atoms.

The magnetic moments are found to be oriented along the $a$ axes in agreement with the experimental findings. The total energy calculated along the $a$ axes is with order 1 and 6 $\mu$Ry lower than the energy calculated along the $c$ and $b$ axes, indicating a very low magneto-crystalline anisotropy for this system. The Curie temperature was estimated to be 250 K using Monte Carlo method implemented in UppASD \cite{uppasd} code.

The Fe hyperfine field B$_{\rm hf}$ was calculated by evaluating the Fermi contact contribution \begin{equation}B_{hf} = \frac{2}{3} \mu_0 \mu_B [\rho^{\uparrow}(0) - \rho^{\downarrow}(0)]\end{equation} where $\rho^{\uparrow}$(0) and $\rho^{\downarrow}$(0) are the spin-up and spin-down densities at the Fe nucleus. Contributions to the field from both core (1s, 2s and 3s) and valence electrons (4s) were calculated.  Calculations were first made with frozen core and at the latest stage with relaxed core for the equilibrium structure. The B$_{\rm hf}$ (core) is negative -12.12 T and B$_{\rm hf}$ (valence) is positive 10.73 T and the calculated Fe magnetic moment is 1.07 $\mu_B$. The experimental value for the moment is found to be 1.4(3) $\mu_B$. It has been found that the B$_{\rm hf}$ (core) is proportional to the magnetic moment \cite{Kamali2011} why it seems appropriate to correct for the magnetic moment mismatch. Thus a corrected total magnetic hyperfine field could be calculated to B$_{\rm hf}$ = -5.13 T, which is in rather good agreement with the experimental saturated value of -8.87 T \cite{Chai2014}.

\section{Conclusions}
\label{}
The magnetic structure and magnetocaloric properties of AlFe$_2$B$_2$ have been studied with X-ray and neutron diffraction, magnetic measurements and electronic structure calculations. The magnetic moments are found to be oriented along the crystallographic $a$-axis from both the neutron diffraction experiments and  the electronic structure calculations. 

The magnetic measurements reveal that a second order magnetic phase transition occurs at 285 K. The magnetic entropy change for this specific sample is -1.3 J/kg K for a change of the magnetic field from 0 to 800 kA/m.

\section*{Acknowledgments}
Financial support from the Swedish Research Council is gratefully acknowledged. E. K. D. Cz.  and L. B. acknowledges SNIC NSC-Triolith resources for computational support.





\bibliographystyle{model1a-num-names}
\bibliography{references}







\end{document}